\documentclass[letter]{ieice}
\usepackage{graphicx}
\field{}
\title{ Agent-Based Proof Design via Lemma Flow Diagram}
\authorlist{
\authorentry{Keehang Kwon}{m}{labelA}
\authorentry{Daeseong Kang}{n}{labelB}
}
\affiliate[labelA]{The author is with Computer Eng., DongA University. email:khkwon@dau.ac.kr}
\affiliate[labelB]{The author is with Electronics Eng., DongA University. email:dskang@dau.ac.kr}

\newcommand{\fol}{\mbox{FOL}}
\newcommand{\folw}{\mbox{FOL$^\Omega$}}

\newcommand{\lfd}{LFD}

\let\Entails\vdash

\def\implKern{\kern-0.11em}

\newcommand{\Rule}[2]
  {\ensuremath{\begin{array}[b]{@{}c@{}}#1\\\hline#2\end{array}}}
\newcommand{\RuleMW}[3]
  {\RuleLW{#1}{\ensuremath{#2}}{#3}}
\newcommand{\RuleM}[3]
  {\RuleL{#1}{\ensuremath{#2}}{#3}}
\newcommand{\RuleTW}[3]
  {\RuleLW{#1}{\textsf{#2}}{#3}}
\newcommand{\RuleT}[3]
  {\RuleL{#1}{\textsf{#2}}{#3}}

\newsavebox{\infRuleBot}\newsavebox{\infRuleLabel}\newsavebox{\infRule}
\newlength{\infRuleWidth}\newlength{\infRuleLblHeight}\newlength{\infRuleCenter}
\newcommand{\RuleXXX}[4]
  {\sbox{\infRuleBot}{\ensuremath{#3}}
   \sbox{\infRuleLabel}{#2}
   \sbox{\infRule}{\Rule{#1}{\usebox{\infRuleBot}}}
   \settowidth{\infRuleWidth}{\usebox{\infRule}}
   \settoheight{\infRuleCenter}{\usebox{\infRuleBot}}
   \settoheight{\infRuleLblHeight}{\usebox{\infRuleLabel}}
   \addtolength{\infRuleCenter}{-0.5\infRuleLabel}
   \addtolength{\infRuleCenter}{-0.4ex}
   #4{\usebox{\infRule}%
      \raisebox{\infRuleCenter}{\usebox{\infRuleLabel}}}}
\newcommand{\RuleLW}[3]
  {\RuleXXX{#1}{#2}{#3}{\makebox}}
\newcommand{\RuleL}[3]
  {\RuleXXX{#1}{#2}{#3}{\makebox[\infRuleWidth][l]}}

\newcommand{\Cut}
  {\RuleTW{\Delta_1\Entails^{m_1} B_1 \ldots \quad \Delta_n\Entails^{m_n} B_n
 \quad  B_1,\ldots,B_n,\Gamma\Entails^{m} C}
  {}
  {\Delta_1,\ldots,\Delta_n, \Gamma\Entails^m C}}

\newcommand{\Cuttwo}
  {\RuleTW{\Delta_1\Entails^{m_1}  \ldots \quad \Delta_n\Entails^{m_n} 
 \quad  B_1^{m_1},\ldots,B_n^{m_n},\Gamma\Entails^m C}
  {}
  {\Delta_1,\ldots,\Delta_n, \Gamma\Entails^m C}}

\newcommand{\oo}{\bot}            
\newcommand{\pp}{\top}

\newcommand{\gneg}{\mbox{\small $\neg$}}                  
\newcommand{\mli}{\hspace{2pt}\mbox{\small $\rightarrow$}\hspace{2pt}}                      
\newcommand{\cla}{\mbox{$\forall$}}      
\newcommand{\cle}{\mbox{$\exists$}}        
\newcommand{\mld}{\hspace{2pt}\mbox{\small $\vee$}\hspace{2pt}}     
\newcommand{\mlc}{\hspace{2pt}\mbox{\small $\wedge$}\hspace{2pt}}   
 
\newcommand{\tlg}{\bot}               
\newcommand{\twg}{\top}               

\newtheorem{theoremm}{Theorem}[section]
\newtheorem{factt}[theoremm]{Fact}
\newtheorem{corollaryy}[theoremm]{Corollary}
\newtheorem{definitionn}[theoremm]{Definition}
\newtheorem{thesiss}[theoremm]{Thesis}
\newtheorem{lemmaa}[theoremm]{Lemma}
\newtheorem{conventionn}[theoremm]{Convention}
\newtheorem{examplee}[theoremm]{Example}
\newtheorem{exercisee}[theoremm]{Exercise}
\newtheorem{remarkk}[theoremm]{Remark}
\newenvironment{definition}{\begin{definitionn} \em}{ \end{definitionn}}

\newenvironment{lemma}{\begin{lemmaa}}{\end{lemmaa}}

\newcounter{itemno}

\newcounter{itemno1}

\newcounter{itemno2}

\newcounter{exno}

\newcounter{defno}

\newcommand{\oprove}{\vdash\kern-.6em\lower.7ex\hbox{$\scriptstyle O$}\,}

\newcommand{\pderivation}{{\cal P}\kern -.1em\hbox{\rm -derivation}}
\newcommand{\pderivationl}{{\cal P}\kern -.1em\hbox{\em -derivation}}
\newcommand{\pderivable}{{\cal P}\kern -.1em\hbox{\rm -derivable}}
\newcommand{\pderivablel}{{\cal P}\kern -.1em\hbox{\em -derivable}}
\newcommand{\pderivations}{{\cal P}\kern -.1em\hbox{\rm -derivations}}
\newcommand{\pderivability}{{\cal P}\kern -.1em\hbox{\rm -derivability}}

\newsavebox{\lpartfig}
\newsavebox{\rpartfig}

\newenvironment{exmple}{
 \begingroup \begin{tabbing} \hspace{2em}\= \hspace{3em}\= \hspace{3em}\=
\hspace{3em}\= \hspace{3em}\= \hspace{3em}\= \kill}{
 \end{tabbing}\endgroup}

\begin{document}
\maketitle
\begin{summary}
  We discuss an agent-based approach to proof design and implementation,
which we call {\it Lemma Flow Diagram}. This approach is based on the multicut
 rule
 with  $shared$ cuts. This approach is modular and  easy to use,
read and automate. Some examples are provided.

\end{summary}
\begin{keywords}
proof design,  flow proof, lemma flow diagram.
\end{keywords}

\section{Introduction}\label{sec:intro}

Proofs typically have a great level of complexity and therefore require
a careful {\it proof design} to reduce their complexity.
Despite its importance,  proof design has received little attention. This is in
contrast to the situation where 
 proof procedures -- natural deduction, resolution and sequent
systems, etc -- are well-studied.

In this paper, we  introduce a new  {\it proof design} tool called {\it Lemma Flow Diagram} (\lfd). 
The most central aspect
of  proof design is a set of well-chosen lemmas.
A set of well-chosen lemmas is important in two aspects:

\begin{itemize}
  
\item It improves the readability of the proof.

\item It improves the search for the proof in automated reasoning.
  
\end{itemize}
\noindent In fact, it is a set of lemmas that  makes proofs 
reative, diverse and interesting.  In contrast, proofs without lemmas 
are rather mechanical, unimaginative and quite awkward.

As we shall see, our design tool is natural and very easy to use,
read and automate. 
This is analogous to the situation where
people have introduced software design tools -- UML, etc --
to design complex software.    \lfd\ is modeled    after 
the interactive computing model  of Computability Logic \cite{Jap0}--\cite{JapCL12} and its distributed variant with agent parameters \cite{kwon19d}.

Our starting point for building lemma-based proofs is the multicut(MC) rule. 
The  MC rule  expresses a form of modus pones -- or lemma-based
reasoning --
which is the key element in compact proof design.

Suppose we have well chosen lemmas $B_1,\ldots,B_n$.
Then the (distributed version of) MC rule  is of the form: \\

\[ \Cut   \]
\noindent Here, $\Gamma, \Delta$ are a set of formulas, $B,C$ are formulas, and
each $m,m_1,\ldots,m_n$ is an agent/machine. That is, each $m_i$ tries to
prove $B_i$ with
respect to $\Delta_i$.

The premise of the above rule is easy to implement by distributedly
processing each sequent in the premise
  but there is some unnaturalness in the above rule. That is, each $B_i$ 
can be ``shared'' between two agents but 
there is no indication of this fact.

A second, alternative  way is to represent the MC rule with ``sharing''.
This rule, which we call {\it the MC' rule}, is shown below:

\[ \Cuttwo   \]
\noindent Here, $\Delta_i$ acts as the knowledgebase of $m_i$ and $B_i^{m_i}$ acts
as a query $B_i$ to $m_i$. 

This MC' rule leads to the diagram in Figure 1.

\begin{center}
\begin{figure}
\includegraphics[width=0.5\textwidth]{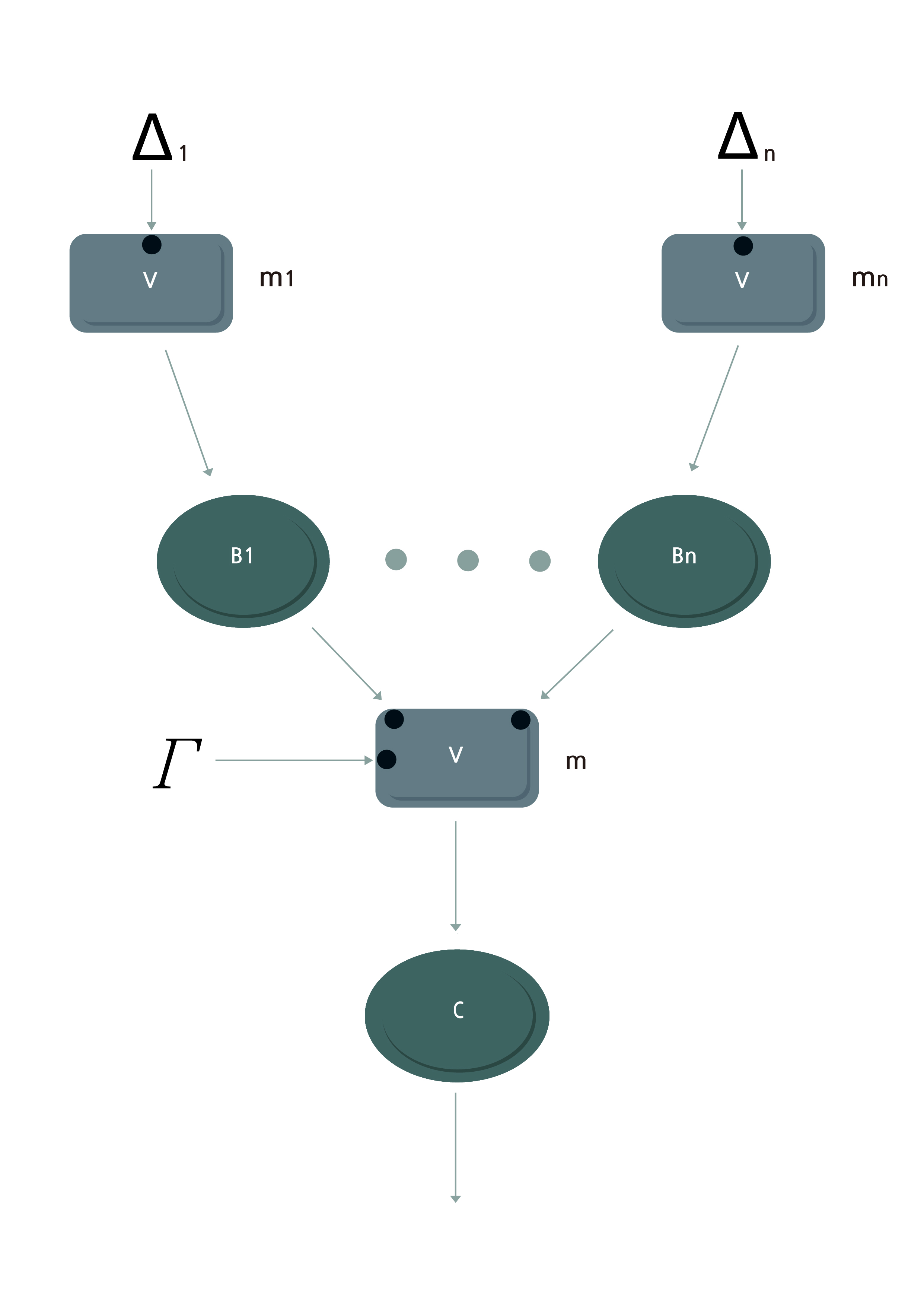}
\caption{\lfd\ for the multicut}
\end{figure}
\end{center}
\noindent
In the diagram, note that each lemma $B_i$ is shared between $m_i$ and $m_{i+1}$.
We call this  diagram \lfd\ with the following features:

\begin{itemize}
\item A regular arrow to box B with circle F means that some unknown identity (e.g. Nature) provides a regular service
$F$ to B. Regular services include $sunny$ or $prime(2)$.

\item A regular  arrow   from box A to box B 
with circle F in the middle means that A provides a querying service F  to B.  

\item Knowledgebase(KB) of an agent is marked with $\bullet$s. In contrast, queries to the agent is marked without $\bullet$s.

\item We assume that the top-level operator associated with an agent is
  $\mld$. That is, an agent with multiple KB services $KB_1,\ldots,KB_m$ and a query services $Q$
should be understood as an agent with a query service 
  $\gneg KB_1 \mld \ldots \mld \gneg KB_m \mld Q$.

\end{itemize}
\noindent
\lfd\ can be seen as a  modular alternative to
the ``flow proof'' which is popular
in mathematical education.


\section{Examples}\label{sec:modules}

As an example of \lfd, we will look at Peano Arithmetic (PA).
PA is formulated with the usual $0,1,+,\times,=$.
This is shown below.

\begin{exmple}
\> $PA$ rules.\\
\>(1)$\cla x\cla y(x+1=y+1 \mli x=y)$.\\
\>(2)$\cla x(x+1\neq 0)$. \\
\>(3)$0+1=1$. \\
\>(4)$\cla x(x+0=x)$.\\
\>(5)$\cla x\cla y(x+(y+1)=(x+y)+1)$.\\
\>(6)$\cla x(x\times 0=0)$.\\
\>(7)$\cla x\cla y(x\times(y+1)=(x+y)+x)$.\\
\>(8)$\cla Q ((Q(0)\mlc \cla x(Q(x)\mli Q(x+1))) \mli \cla x Q(x))$.\\
\end{exmple}
\noindent Rule (8) is called the {\it Induction} rule.

\begin{center}
\begin{figure}
\includegraphics[width=0.5\textwidth]{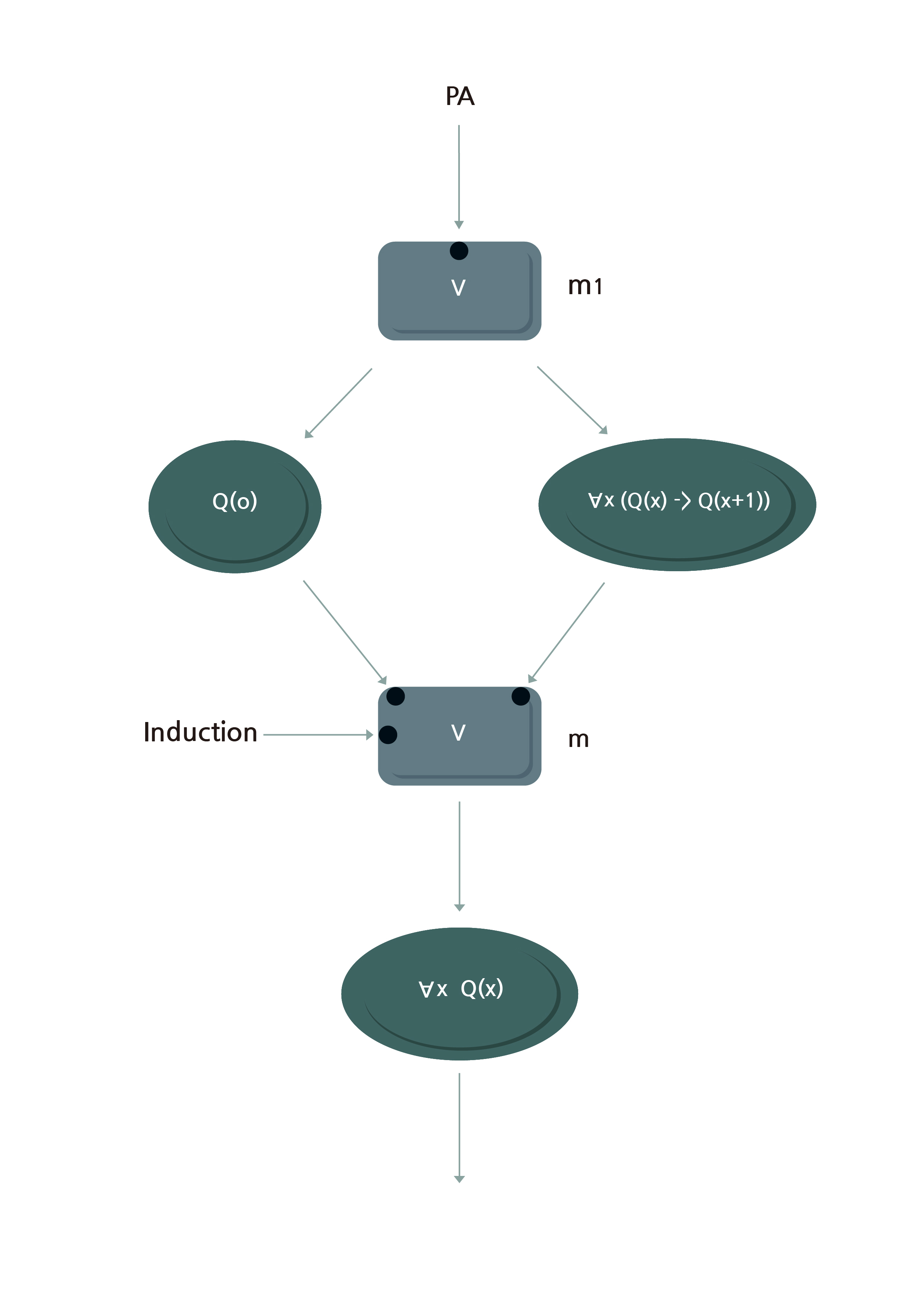}
\caption{\lfd\ for the example}
\end{figure}
\end{center}

Now we want to prove the following:

\[ \cla x (x+1=1+x) \]

The proof requires mathematical induction with $Q(x)$ being $(x+1=1+x)$.
The base case $Q(0)$ which is $(0+1=1+0)$ follows from axiom 3 and 4.
The induction step follows from induction hypothesis and axiom 5.

We display the \lfd\ of the above example in Figure 2.
This diagram is straightforward to understand.

\section{Distributed  first-order logic}\label{s2tb}

To represent the `querying' formulas in the MC' rule, we need a new representation language, because they are not allowed in  classical logic.
For this reason, we introduce \folw\ which provides an elegant solution
in this regard.

\folw\ is based on the game semantics. 
For this reason, we review the  the game-semantical meanings of first-order logic\cite{Japfi}.

First of all,  classical propositions  are viewed as special, {\em elementary} sorts of games that have no moves and are automatically won by the machine if true, and lost if false. For example, the proposition $prime(5)$ is true and is automatically won by the machine.

There are two special atoms $\twg$ and $\tlg$. $\twg$ is always true, and $\tlg$ is always false. 

Negation $\gneg$ is a role-switch operation: $\gneg A$ is obtained from $A$ by turning $\pp$'s  moves and wins into $\oo$'s  moves and wins, and vice versa. For example, if {\em Chess} means the game of chess from the  view of the black player, then $\gneg${\em Chess} is the same game from the  of view of the white player.

The parallel operations deal with  concurrent computations. Playing $A\mlc B$ or $A\mld B$ means playing, in parallel, the two games $A$ and $B$. In $A\mlc B$, $\pp$ is considered the winner if it wins in both of the components, while in $A\mld B$ it is sufficient to win in one of the components. 

The reduction operation $A\mli B$ is understood as $\gneg  A \mld B$.

The blind quantifiers $\cla x A$ and $\cle x A$  model information hiding
  with $\cle x$  meaning $\pp$'s choices, and $\cla x$ 
meaning choices by $\oo$. In both cases, the value of $x$ is unknown or invisible
to the opponent.

Following the ideas in \cite{kwon19d},  we introduce $\folw$, a
slight extension to FOL with agent parameters.
Let $F$ be a formula in first-order logic.
We introduce a new {\it env-annotated} formula $F^\omega$ which reads as `play $F$ against an agent $\omega$.
For  $F^\omega$, we say
$\omega$ is the {\it matching} environment of $F$.

In introducing agent parameters to a formula $F$, 
some formulas turn out to be difficult to process. `Environment-switching' formulas are
such examples. In them, the machine initially provides a service $F$ against agent $w$ and then switches
to provide another service to  another agent $u$ inside $F$. This kind of formulas are  difficult to
process. This leads to the following definition: 

\begin{definition} 
  The class of $\folw$-formulas
is defined as the smallest set of expressions such that
(a) For any $\fol$-formula $F$ and any agent $\omega$, $F^\omega $  is in it and, (b) if $H$ and $J$ are in it, then so are 
$\gneg H, H\mlc J, H\mld J, H\mli J$.
\end{definition}

 Given a  $\folw$-formula $H$, the diagram of $H$ can be easily
 obtained. 

\begin{definition}
\noindent Given a  $\folw$-formula $H$, the diagram of $H$  -- denoted by
$diagram(H)$ -- is obtained by the rules in Figure 3. 
There, $F$ is a $\fol$-formula and $H,J$ are $\folw$-formulas.
Further, $A$ is the current machine, $W$ is an environment.
\end{definition}
\noindent 

\begin{center}
\begin{figure}
\includegraphics[width=0.5\textwidth]{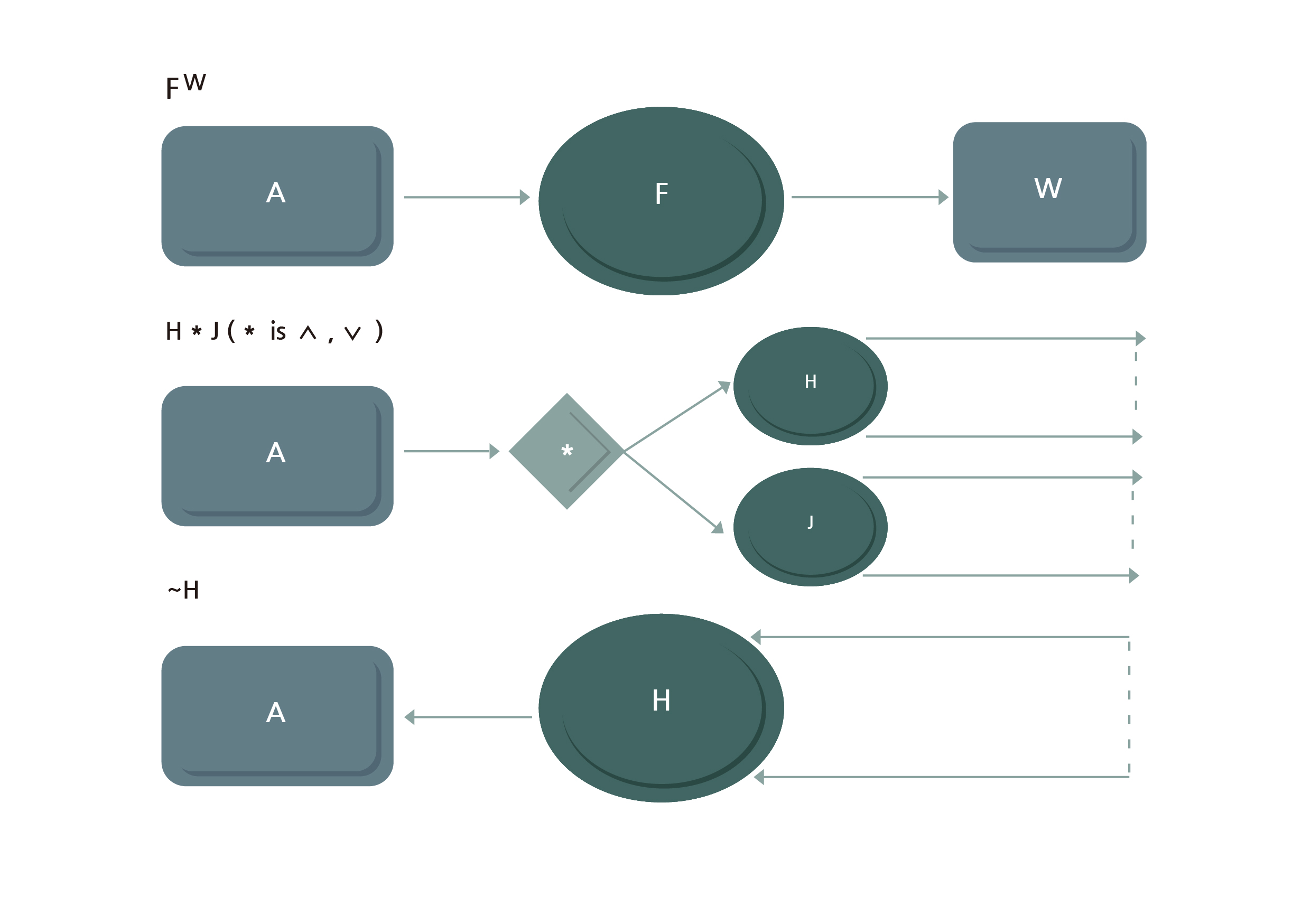}
\caption{\folw-formulas to diagrams} 
\end{figure}
\end{center}

\section{An Example: Encoding the Peano Axiom}\label{sec:modules}

The diagram in Section 2 corresponds to  the following \folw. The
(KB of) agent $m_1$ is shown below.

\begin{exmple}
\> $agent\ m_1$.\\
\>$\cla x\cla y(x+1=y+1 \mli x=y)$.\\
\>$\cla x(x+1\neq 0)$. \\
\>$0+1=1$. \\
\>$\cla x(x+0=x)$.\\
\>$\cla x\cla y(x+(y+1)=(x+y)+1)$.\\
\>$\cla x(x\times 0=0)$.\\
\>$\cla x\cla y(x\times(y+1)=(x+y)+x)$.\\
\end{exmple}

Our language  permits  `querying knowledge' of the form
$Q^\omega$ in KB. This requires the current machine to
invoke the query $Q$ to the agent $\omega$. 
Now let us consider the $m$ agent which handles mathematical induction.
It contains the  induction rule and 
two querying  knowledges  $Q(0)^{m_1}$ and
 $(\cla x(Q(x)\mli Q(x+1)))^{m_1}$.
Note that the induction rule is technically not first-order but it does not affect
our main argument.

\begin{exmple}
 $agent\ m$.\\
 $Q(0)^{m_1}$.\\
 $(\cla x(Q(x)\mli Q(x+1)))^{m_1}$.\\
 $\cla Q ((Q(0)\mlc \cla x(Q(x)\mli Q(x+1))) \mli \cla x Q(x))$.

\end{exmple}

Last, the theorem to prove corresponds to the query 
?-   $\cla x (x+1=1+x)$ with respect to the $m$ agent.

It can be easily observed that the above code can be automated. This
leads to the notion of 
{\it Lemma Flow Theorem Proving}.  
Now, solving the above goal  has the  following strategy: first 
solve both the query $Q(0)^{m_1}$ and $(\cla x(Q(x)\mli Q(x+1)))^{m_1}$
with respect to $m_1$ and then solve $\cla x (x+1=1+x)$ with respect to
$m$.
 Note that all three attempts must succeed for $\cla x (x+1=1+x)$ to be
true.

\section{Conclusion}\label{sec:conc}

In this paper, we have considered a proof design tool called \lfd\ and a distributed first-order logic
\folw. 
 Our diagram  allows modular, distributed proof design. It 
 is therefore useful for reading/writing/automating proofs.
 \folw\ has a simple syntax and semantics and is well-suited to
 representing distributed KB \cite{Loke}.

\section{Acknowledgements}

This work  was supported by DongA University Research Fund.




\bibliographystyle{plain}

\end{document}